\documentclass[10pt,twoside,twocolumn,letterpaper]{IEEEtran}
\pdfoutput=1
\usepackage{graphicx,cite,amsfonts,epsfig,amssymb,amsmath}

\title{Channel-Adapted Quantum Error Correction for the Amplitude Damping Channel}
\author{Andrew~S.~Fletcher,~Peter~W.~Shor,~and~Moe~Z.~Win~\IEEEmembership{Fellow,~IEEE}%
\thanks{This paper is based on a thesis submitted in partial fulfillment of the requirements for the degree of Doctor of Philosophy in the Department of Electrical Engineering and Computer Science at the Massachusetts Institute of Technology in June, 2007.}%
\thanks{A.S.F. would like to thank the Department of the Air Force, who sponsored this work under AF Contract \#FA8721-05-C-0002.   All authors thank the National Science Foundation for support through grant CCF-0431787.  Opinions, interpretations, recommendations and conclusions are those of the authors and are not necessarily endorsed by the United States Government.}}

\newcommand{\E}{\mathcal{E}}
\newcommand{\R}{\mathcal{R}}

\input{Qcircuit}

\begin{document}

\maketitle
\IEEEpeerreviewmaketitle

\begin{abstract}

We consider error correction procedures designed specifically for the amplitude damping channel.  We analyze amplitude damping errors in the stabilizer formalism.  This analysis allows a generalization of the $[4,1]$ `approximate' amplitude damping code of \cite{LeuNieChuYam:97}.  We present this generalization as a class of $[2(M+1),M]$ codes for $M\geq 1$ and present quantum circuits for encoding and recovery operations.  We also present a $[7,3]$ amplitude damping code based on the classical Hamming code.  All of these are stabilizer codes whose encoding and recovery operations can be completely described with Clifford group operations.  Finally, we describe optimization options in which recovery operations may be further adapted according to the damping probability $\gamma$.

\end{abstract}

\section{Introduction}

In the most common treatments, quantum error correction (QEC) is developed for a very generic error model.  An arbitrary error on a single qubit is correctable if both Pauli $X$ and $Z$ are correctable on that qubit; the continuum of quantum errors are thus reduced into a simple, discrete set.  Using this approach, we may design error correction procedures that apply to a wide variety of quantum noise processes -- the channel need only be well approximated by independent qubit errors.

The general application of standard QEC comes with a price in efficiency.  Quantum error correcting codes require a large number of redundant qubits; for short block lengths, generic codes are limited to low rates.  While robust to arbitrary qubit errors, both error correction performance and efficiency can be improved by adapting the encoding and recovery operations to the physical noise process.  Such adaptation is reasonable since, for any particular device, the noise will have a structure governed by the physical coupling of the system and the environment.  Intuitively, we should be able to engineer improved error correction by careful adaptation of both the encoding and recovery operations.

The concept of channel-adapted error correction is not new: early work labeled `approximate' quantum error correction was presented in \cite{LeuNieChuYam:97}.  Much recent progress has been due to optimization efforts \cite{FleShoWin:07,Fle:07,FleShoWin:J07a,KosLid:06,ReiWer:05,YamHarTsu:05}.  In each case, rather than correcting for arbitrary single qubit errors, the error recovery scheme was adapted to a model for the noise, with the goal to maximize the fidelity of the operation.  In \cite{FleShoWin:07}, a semidefinite program (SDP) was used to maximize the entanglement fidelity, given a fixed encoding and channel model.  In \cite{KosLid:06} and \cite{ReiWer:05}, encodings and decodings were iteratively improved using the performance criteria of ensemble average fidelity and entanglement fidelity, respectively.  A sub-optimal method for minimum fidelity, using an SDP, was proposed in \cite{YamHarTsu:05}.  An analytical approach to channel-adapted recovery based on the pretty-good measurement and the average entanglement fidelity was derived in \cite{BarKni:02}.  The main point of each scheme was to improve error corrective procedures by adapting to the physical noise process.

The optimization efforts cited above detail mathematical and algorithmic tools with general application.  That is to say, given any model for the noise process and an appropriately short code we can apply optimal\cite{FleShoWin:07,KosLid:06} and structured near-optimal \cite{FleShoWin:J07a} algorithms to provide channel-adapted encoding and recovery operations.

It is important to note that the aforementioned tools are not, in themselves, complete solutions to the problem of channel-adapted QEC.  When designing an error correction procedure, there is more to consider than whether an encoding or a recovery is physically legitimate.  This motivated our exploration of near-optimal recovery operations\cite{FleShoWin:J07a}, where we imposed a projective syndrome measurement constraint on recovery operations.  Even given such a constraint, to implement channel-adapted QEC efficiently we need to design encoding and decoding procedures with sufficiently simple  structure to allow efficient implementation.  Furthermore, while the optimization routines focus on the entanglement fidelity and ensemble average fidelity due to their linearity, we should still like to understand the minimum fidelity, or worst case performance.

To explore these issues in greater depth, we must consider channel-adapted QEC for a specific channel model.  We examine the amplitude damping channel, denoted $\E_a$, given by the operation elements
\begin{equation}\label{eq:ampdamp}
E_0=\left [ \begin{array}{ccc} 1 & 0 \\ 0 &\sqrt{1-\gamma} \end{array} \right ]\hspace{.5 cm} \textrm{and} \hspace{.5 cm}
E_1=\left [ \begin{array}{ccc} 0 & \sqrt{\gamma} \\ 0 & 0 \end{array} \right ].
\end{equation}
Amplitude damping is a logical choice for several reasons.  First of all, it has a useful physical interpretation: the parameter $\gamma$ indicates the probability of decaying from state $\ket{1}$ to $\ket{0}$ (\emph{i.e.}~the probability of losing a photon).  Second, amplitude damping cannot be written with scaled Pauli matrices as the operator elements; thus Theorem 1 from \cite{Fle:07} does not apply and the optimal recovery operation does not have a near-trivial form.  Finally, due to its structure, the amplitude damping channel can still be described with the stabilizer formalism, greatly aiding analysis.

\section{Qualitative analysis of channel-adapted QER for approximate [4,1] code}\label{sec:qualitative analysis 4,1}

\begin{table}[bt]
\begin{center}
  \begin{tabular}{|c|c|}
    \hline
    $R_1$ & $\ket{0_L}(\alpha \bra{0000} +\beta\bra{1111})+\ket{1_L}(\frac{1}{\sqrt{2}}\bra{0011}+\frac{1}{\sqrt{2}}\bra{1100})$\\
    $R_2$ & $\ket{0_L}(\beta \bra{0000} -\alpha\bra{1111})+\ket{1_L}(\frac{1}{\sqrt{2}}\bra{0011}-\frac{1}{\sqrt{2}}\bra{1100})$\\
    $R_3$ & $\ket{0_L}\bra{0111}+\ket{1_L}\bra{0100}$\\
    $R_4$ & $\ket{0_L}\bra{1011}+\ket{1_L}\bra{1000}$\\
    $R_5$ & $\ket{0_L}\bra{1101}+\ket{1_L}\bra{0001}$\\
    $R_6$ & $\ket{0_L}\bra{1110}+\ket{1_L}\bra{0010}$\\
    $R_7$ & $\ket{0_L}\bra{1001}$\\
    $R_8$ & $\ket{0_L}\bra{1010}$\\
    $R_9$ & $\ket{0_L}\bra{0101}$\\
    $R_{10} $& $\ket{0_L}\bra{0110}$\\
    \hline
  \end{tabular}
  \end{center}
  \caption[Optimal QER operator elements for the 4 qubit code.]{Optimal QER operator elements for the [4,1] code.  Operators $R_1$ and $R_2$ correspond to the ``no dampings'' term $E_0^{\otimes 5}$ where $\alpha$ and $\beta$ depend on $\gamma$.  $R_3-R_6$ correct first order dampings.  $R_7-R_{10}$ partially correct some second order dampings, though as only $\ket{0_L}$ is returned in these cases superposition is not preserved.}\label{tab:4 qubit recovery}
\end{table}

We begin with a qualitative understanding of the $[4,1]$ `approximate' code of \cite{LeuNieChuYam:97} and its optimal channel-adapted recovery operation.  The logical codewords are given by
\begin{eqnarray}\label{eq:4qubitcode_v2}
\ket{0_L}&=& \frac{1}{\sqrt{2}}(\ket{0000}+\ket{1111})\\
\ket{1_L}&=& \frac{1}{\sqrt{2}}(\ket{0011}+\ket{1100})\label{eq:4qubitcode1_v2}.
\end{eqnarray}
Consider the optimal channel-adapted recovery for the [4,1] `approximate' code of \cite{LeuNieChuYam:97} computed with the semidefinite program as presented in \cite{FleShoWin:07}.  This is an example of a channel-adapted code, designed specifically for the amplitude damping channel rather than arbitrary qubit errors.  Its initial publication demonstrated the utility of channel-adaptation (though without using such a term) for duplicating the performance of standard quantum codes with both a shorter block length and while achieving a higher rate.

In \cite{LeuNieChuYam:97}, the authors proposed a recovery (decoding) circuit and demonstrated its strong performance in minimum fidelity.  It is interesting to note that the recovery operation (described in quantum circuit form in Fig.~2 of \cite{LeuNieChuYam:97}) is not a projective syndrome measurement followed by a unitary rotation as is standard for generic codes;  instead it is a $\gamma$-dependent operation which includes a generalized (POVM) measurement.  In contrast, the optimal recovery (in terms of entanglement fidelity) obtained via convex optimization \emph{does} conform to such a structure.
The optimal recovery operation is given in Table \ref{tab:4 qubit recovery}.  We will analyze each of the operator elements in turn.  For clarity of presentation, we begin with first and second order damping errors and then we turn our attention to the recovery from the `no damping' term.

\subsection{Recovery for first and second order damping errors}

Neither $E_0$ nor $E_1$ in (\ref{eq:ampdamp}) is a scaled unitary matrix, but we may understand the channel by considering $E_1$ the `error' event.  Let us denote a first order damping error as $E_1^{(k)}$, which consists of the qubit operator $E_1$ on the $k^{th}$ qubit and the identity elsewhere.  Consider now the effect of $E_1^{(1)}$ on the codewords of the $[4,1]$ code:
\begin{eqnarray}
  E_1\otimes I^{\otimes 3} \ket{0_L}=\sqrt{\gamma}\ket{0111},\\
  E_1\otimes I^{\otimes 3} \ket{1_L}=\sqrt{\gamma}\ket{0100}.
\end{eqnarray}
We see that the code subspace is perturbed onto an orthogonal subspace spanned by $\{\ket{0111},\ket{0100}\}$.  $R_3$ projects onto this syndrome subspace and recovers appropriately into the logical codewords.  Recovery operators $R_4$, $R_5$, and $R_6$ similarly correct damping errors on the second, third, and fourth qubits.  Notice that the first order damping errors move the information into mutually orthogonal subspaces.  It is therefore not hard to see that the set of errors $\{I^{\otimes 4},E_1^{(k)}\}_{k=1}^4$ satisfy the error correcting conditions for the $[4,1]$ code.  (That the $[4,1]$ code satisfies the error correcting conditions for damping errors was pointed out in \cite{Got:97}.)

Consider now the subspace spanned by $\{\ket{1010},\ket{0101},\ket{0110},\ket{1001}\}$.  By examining the logical codewords in (\ref{eq:4qubitcode_v2}) and (\ref{eq:4qubitcode1_v2}), we see that this subspace can only be reached by multiple damping errors.  Unfortunately, in such a case we lose the logical superpositions as only $\ket{0_L}$ is perturbed into this subspace.  Consider, for example the two damping error $E_1^{(1)}E_1^{(3)}$.  We see that
\begin{eqnarray}
  E_1^{(1)}E_1^{(3)}\ket{0_L}&=&\gamma\ket{0101},\\
  E_1^{(1)}E_1^{(3)}\ket{1_L}&=&0.
\end{eqnarray}
While we cannot fully recover from such an error, we recognize that these higher order errors occur with probability $\gamma^2$.  Furthermore, we see that operator elements $R_7-R_{10}$ do recover the $\ket{0_L}$ portion of the input information.   This contributes a small amount to the overall entanglement fidelity, though would obviously not help the minimum fidelity case.  Indeed, $R_7-R_{10}$ do not contribute to maintaining the fidelity of an input $\ket{1_L}$ state.

We should also note that only a subset of all second order dampings are partially correctable as above.  We reach the syndrome subspaces from $R_7-R_{10}$ only when a qubit from the first pair and a qubit from the second pair is damped, allowing the $\ket{0_L}$ state to be recovered.  If both the first and second qubits (or both the third and fourth qubits) are damped, the resulting states are no longer orthogonal to the code subspace.  In fact, these are the only errors that will cause a logical bit flip, recovering $\ket{0_L}$ as $\ket{1_L}$ and vice versa.

\subsection{Recovery from the distortion of the `no damping' case}\label{sec:4qubit no damping}

We turn now to the recovery operators $R_1$ and $R_2$.  Together these project onto the syndrome subspace with basis vectors $\{\ket{0000},\ket{1111},\ket{1100},\ket{0011}\}$ which includes the entire code subspace.  We just saw that $I^{\otimes 4}$ together with single qubit dampings are correctable, but $\E_{a}^{\otimes 4}$ does not have an operator element proportional to $I^{\otimes 4}$.  Instead, the `no dampings' term is given by $E_0^{\otimes 4}$ which depends on the damping parameter $\gamma$.  Indeed, consider the effect of the no damping term on the logical code words:
\begin{eqnarray}
  E_0^{\otimes 4}\ket{0_L}&=&\frac{1}{\sqrt{2}}(\ket{0000}+(1-\gamma)^2\ket{1111})\\
  E_0^{\otimes 4}\ket{1_L}&=&\frac{1-\gamma}{\sqrt{2}}(\ket{1100}+\ket{0011}).
\end{eqnarray}

A standard recovery operation projects onto the code subspace.  Consider the effect of such a recovery on an arbitrary input state $a\ket{0_L}+b\ket{1_L}$. The resulting (un-normalized) state is
\begin{equation}
  a(1-\gamma+\frac{\gamma^2}{2})\ket{0_L}+b(1-\gamma)\ket{1_L}.
\end{equation}
The extra term $\frac{\gamma^2}{2}$ distorts the state from the original input.  While this distortion is small as $\gamma\rightarrow 0$, both the original recovery operation proposed in \cite{LeuNieChuYam:97} and the optimal recovery seek to reduce this distortion by use of a $\gamma$-dependent operation.  We analyze the optimal recovery operation for this term and compare its efficacy with the simpler projection.

We see that $R_1$ projects onto a perturbed version of the codespace with basis vectors $\{(\alpha\ket{0000}+\beta\ket{1111}),(\frac{1}{\sqrt{2}}\ket{0011}+\frac{1}{\sqrt{2}}\ket{1100})\}$ where $\alpha$ and $\beta$ are chosen to maximize the entanglement fidelity.  We can use any of the numerical techniques of \cite{FleShoWin:07,FleShoWin:J07a} to compute good values for $\alpha$ and $\beta$, but we would like an intuitive understanding as well.  $\alpha$ and $\beta$ (where $|\beta|=\sqrt{1-|\alpha|^2}$) adjust the syndrome measurement $P_1$ so that it is no longer $\ket{0_L}\bra{0_L}+\ket{1_L}\bra{1_L}$, the projector onto the code subspace.  If we choose them so that $\bra{0_L}P_1\ket{0_L}=\bra{1_L}P_1\ket{1_L}$ then we will perfectly recover the original state when syndrome $P_1$ is detected for the no damping case.  If syndrome $P_2$ is detected, the no damping state will be distorted, but for small $\gamma$, the second syndrome is a relatively rare occurrence.  It could even be used as a classical indicator for a greater level of distortion.

\begin{figure}
  \begin{center}
    \includegraphics[width=\columnwidth]{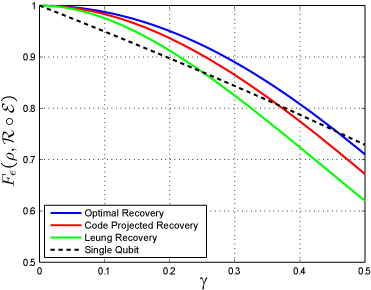}
    \caption[Optimal vs.~code projection recovery operations for the four qubit code.]{Optimal vs.~code projection recovery operations for the [4,1] code.  We compare the entanglement fidelity for the optimal recovery operation and the recovery that includes a projection onto the code subspace.  For comparison, we also include the original recovery operation proposed in \cite{LeuNieChuYam:97} and the baseline performance of a single qubit.  While the optimal recovery outperforms the code projector recovery, the performance gain is likely small compared to the cost of implementing the optimal.}\label{fig:AmpDamp4_stab_rec}
  \end{center}
\end{figure}

We can see in Fig.~\ref{fig:AmpDamp4_stab_rec} that the benefit of the optimal recovery operation is small, especially as $\gamma\rightarrow 0$, though not negligible.  Furthermore, the standard projection onto the code space is a simple operation while the optimal recovery is both $\gamma$-dependent and relatively complex to implement.  For this reason, it is likely preferable to implement the more straightforward code projection, which still reaps most of the benefits of channel-adaptation.

%$|\alpha|^2+|\beta|^2$ must be 1, and for small $\gamma$ we anticipate that both will be close to $1/\sqrt{2}$.  We can accomplish both of these by writing $\alpha=\cos(\pi/4+\theta)$ and $\beta=\sin(\pi/4+\theta)$  and choosing $\theta$ so that the probability of measuring the $\ket{0_L}$ state is equal to the probability of measuring $\ket{1_L}$:
%\begin{eqnarray}
%  1-\gamma&=&\frac{\cos(\pi/4+\theta)+\sin(\pi/4+\theta)(1-\gamma)^2}{\sqrt{2}}\\
%  &\approx&\frac{1}{2}(1-\theta-\frac{\theta^2}{2}+(1+\theta-\frac{\theta^2}{2})(1-\gamma)^2)
%\end{eqnarray}

\section{Amplitude damping errors in the stabilizer formalism}

%We first interpret the recovery in terms of the code words and then in terms of the code stabilizers.  We will see that we can understand both the dominant errors and the recovery operation in terms of stabilizer operations.  The stabilizer interpretation permits a simple generalization for higher rate amplitude damping codes and recovery operations.  In particular, we define two classes of amplitude damping-adapted error correcting codes that can be derived and understood with a simple stabilizer structure.

We turn our attention to the stabilizer formalism\cite{Got:96,Got:97}, to demonstrate its utility for interpreting the amplitude damping channel.  In particular, understanding amplitude damping in terms of stabilizers allows a generalization of the $[4,1]$ code for higher rates.  The stabilizer formalism provides an extremely useful and compact description for quantum error correcting codes.  Code descriptions, syndrome measurements, and recovery operations can be understood by considering the $n-k$ generators of an $[n,k]$ stabilizer code.  In standard practice, Pauli group errors are considered and if $\{X_i,Y_i,Z_i\}_{i=1}^n$ errors can be corrected, we know we can correct an arbitrary error on one of the qubits since the Pauli operators are a basis for single qubit operators.

Let's consider the $[4,1]$ code in terms of its stabilizer group $G=\langle XXXX,ZZII,$\\$IIZZ\rangle$.  We can choose the logical Pauli operators $\bar{X}=XXII$ and $\bar{Z}=ZIZI$ to specify the codewords in (\ref{eq:4qubitcode_v2}) and (\ref{eq:4qubitcode1_v2}).  We saw in Sec.~\ref{sec:qualitative analysis 4,1} that $E_1^{(i)}$ damping errors together with $I^{\otimes4}$ are correctable errors.  Since each of these errors is a linear combination of Pauli group members:
\begin{equation}
  E_1^{(i)}=\frac{\sqrt{\gamma}}{2}(X_i+iY_i),
\end{equation}
we might presume that $\{I,X_i,Y_i\}_{i=1}^4$ are a set of correctable operations and the desired recovery follows the standard stabilizer syndrome measurement structure.  This is not the case.  Consider that the operator $X_1X_2$ (or equivalently $XXII$) is in the normalizer $N(G)$ of the code stabilizer, and thus $\{X_1,X_2\}$ are not a correctable set of errors.

How, then, can the $[4,1]$ code correct errors of the form $X_i+iY_i$?  Instead of projecting onto the stabilizer subspaces and correcting $X_i$ and $Y_i$ separately, we take advantage of the fact that the errors happen in superposition and project accordingly.  As we saw, $X_i+iY_i$ and $X_j+iY_j$ project into orthogonal subspaces when $i\neq j$ and we can recover accordingly.  In fact, the correct syndrome structures can also be described in terms of stabilizers; understanding these syndromes enables design and analysis of other amplitude damping codes.

Let $G=\langle g_1,\ldots,g_{n-k}\rangle$ be the generators for an $[n,k]$ stabilizer code.  We wish to define the generators for the subspace resulting from a damping error $X_i+iY_i$ on the $i^{th}$ qubit.  First, we should note that we can always write the generators of $G$ so that at most one generator commutes with $X_i$ and anti-commutes with $Y_i$ (corresponding to a generator with an $X$ on the $i^{th}$ qubit), at most one generator that anti-commutes with both $X_i$ and $Y_i$ (corresponding to a generator with an $Z$ on the $i^{th}$ qubit), and all other generators commute with both operators.  Let $\ket{\psi}\in C(G)$ be an arbitrary state in the subspace stabilized by $G$.  If $g\in G$ such that $[g,X_i]=[g,Y_i]=0$, then
\begin{equation}
  (X_i+iY_i)\ket{\psi}=(X_i+iY_i)g\ket{\psi}=g(X_i+iY_i)\ket{\psi}.
\end{equation}
From this we see that the $i^{th}$ damped subspace is stabilized by the commuting generators of $G$.  Now consider an element of $G$ that anti-commutes with $X_i$ and $Y_i$.  Then
\begin{equation}
  (X_i+iY_i)\ket{\psi}=(X_i+iY_i)g\ket{\psi}=-g(X_i+iY_i)\ket{\psi},
\end{equation}
so $-g$ is a stabilizer of the $i^{th}$ damped subspace.  Finally, consider a $g$ which commutes with $X_i$ but anti-commutes with $Y_i$:
\begin{equation}
  (X_i+iY_i)\ket{\psi}=(X_i+iY_i)g\ket{\psi}=g(X_i-iY_i)\ket{\psi}.
\end{equation}
We see that neither $g$ nor $-g$ is a stabilizer for the subspace.  It is, however, not hard to see that $Z_i$ is a generator:
\begin{equation}
  Z_i(X_i+iY_i)\ket{\psi}=(iY_i-i^2X_i)\ket{\psi}=(X_i+iY_i)\ket{\psi}.
\end{equation}
In this manner, given any code stabilizer $G$, we can construct the stabilizer for each of the damped subspaces.

\begin{table}
\begin{center}
\begin{tabular}{c@{\hspace{30pt}}c}
\begin{tabular}{c}
    $1^{st}$ subspace\\
    \hline
  \begin{tabular}{c@{}c@{}c@{}c@{}c}
    -&$Z$&$Z$&$I$&$I$\\
    &$I$&$I$&$Z$&$Z$\\
    &$Z$&$I$&$I$&$I$
  \end{tabular}
\end{tabular}
  &
\begin{tabular}{c}
    $2^{nd}$ subspace\\
    \hline
   \begin{tabular}{c@{}c@{}c@{}c@{}c}
   -&$Z$&$Z$&$I$&$I$\\
    &$I$&$I$&$Z$&$Z$\\
    &$I$&$Z$&$I$&$I$
  \end{tabular}
\end{tabular}
  \\
  \\
\begin{tabular}{c}
    $3^{rd}$ subspace\\
    \hline
  \begin{tabular}{c@{}c@{}c@{}c@{}c}
    &$Z$&$Z$&$I$&$I$\\
    -&$I$&$I$&$Z$&$Z$\\
    &$I$&$I$&$Z$&$I$
  \end{tabular}
\end{tabular}
  &
\begin{tabular}{c}
    $4^{th}$ subspace\\
    \hline
  \begin{tabular}{c@{}c@{}c@{}c@{}c}
    &$Z$&$Z$&$I$&$I$\\
    -&$I$&$I$&$Z$&$Z$\\
    &$I$&$I$&$I$&$Z$
  \end{tabular}
\end{tabular}
\end{tabular}
  \end{center}
  \caption[Stabilizers for each of the damped subspaces of the four qubit code.]{Stabilizers for each of the damped subspaces of the $[4,1]$ code.}\label{tab:4 qubit damped subspaces}
\end{table}

Consider now the stabilizer description of each of the damped subspaces for the $[4,1]$ code.  These are given in Table \ref{tab:4 qubit damped subspaces}.  Recall that two stabilizer subspaces are orthogonal if and only if there is an element $g$ that stabilizes one subspace while $-g$ stabilizes the other.  It is easy to see that each of these subspaces is orthogonal to the code subspace, as either $-ZZII$ or $-IIZZ$ is included.  It is equally easy to see that the first and second subspaces are orthogonal to the third and fourth.  To see that the first and second subspaces are orthogonal, note that $-IZII$ stabilizes the first subspace, while $IZII$ stabilizes the second.  Equivalently, $-IIZI$ stabilizes the fourth subspace, thus making it orthogonal to the third.

We can now understand the optimal recovery operation in terms of the code stabilizers.  Consider measuring $ZZII$ and $IIZZ$.  If the result is $(+1,+1)$ then we conclude that no damping has occurred and perform the non-stabilizer operations of $R_1$ and $R_2$ to minimize distortion.  If we measure $(-1,+1)$ we know that either the first or the second qubit was damped.  We can distinguish by measuring $ZIII$, with $+1$ indicating a damping on the first qubit and $-1$ a damping on the second.  If our first syndrome is $(+1,-1)$, we can distinguish between dampings on the third and fourth by measuring $IIZI$.
If our first syndrome yields $(-1,-1)$ we conclude that multiple dampings occurred.  We could simply return an error, or we can do the partial corrections of $R_7-R_{10}$ by further measuring both $ZIII$ and $IIZI$.  It is worth pointing out a feature of the stabilizer analysis highlighted by this multiple dampings case.  Each of the damping subspaces from Table \ref{tab:4 qubit damped subspaces} has three stabilizers and thus encodes a 2 dimensional subspace.  Consider applying $E_1^{(1)}$ to the third damped subspace, equivalent to damping errors on qubits 1 and 3.  Note that there is no generator with an $X$ in the first qubit; the resulting subspace is stabilized by
\begin{equation}
  \langle -ZZII,-IIZZ,IIZI,ZIII \rangle.
\end{equation}
As this has four independent generators, the resulting subspace has dimension 1.  We saw this in the previous section, where for multiple dampings the recovery operation does not preserve logical superpositions but collapses to the $\ket{0_L}$ state.

Stabilizer descriptions for amplitude damping-adapted codes are quite advantageous.  Just as in the case of standard quantum codes, the compact description facilitates analysis and aids design.  While the recovery operations for the amplitude damping codes are not quite as neatly described as the standard stabilizer recovery, the stabilizer formalism facilitates the description.  Furthermore, by considering stabilizer descriptions of the $[4,1]$ code and its recovery operation, we may design other channel-adapted amplitude damping codes.  We will rely on stabilizers throughout the remainder of the paper.

\section{Generalization of the [4,1] code for higher rates}\label{sec:generalized 41}

The stabilizer analysis for the $[4,1]$ code provides a ready means to generalize for higher rate code.  Consider the two codes given in Table \ref{tab:six-two and eight-three} (A).  Each of these is an obvious extension of the $[4,1]$ code, but with a higher rate.  Indeed the general structure can be extended as far as desired generating an $[2(M+1),M]$ code for all positive integers $M$.  We can thus generate a code with rate arbitrarily close to $1/2$.

While the codes presented in Table \ref{tab:six-two and eight-three} (A) have an obvious pattern related to the $[4,1]$ code, we will find it more convenient to consider the stabilizer in standard form as given in Table \ref{tab:six-two and eight-three} (B).  The standard form, including the choice of $\bar{X}_i$ and $\bar{Z}_i$, provides a systematic means to write the encoding circuit.  The change is achieved through a reordering of the qubits which, due to the symmetry of the channel, has no effect on the error correction properties.

Let's consider the form of the $M+2$ stabilizer group generators.  Just as with the $[4,1]$ code, the first generator has an $X$ on every qubit.  The physical qubits are grouped into $M+1$ pairs; for each pair $(i,j)$ there is a generator $Z_iZ_j$.

The structure of the stabilizers makes it easy to see that $\{I^{\otimes 2(M+1)},E_1^{(k)}\}_{k=1}^{2(M+1)}$ satisfy the error correcting conditions for the $[2(M+1),M]$ code.  To see this, we will show that the damped subspaces are mutually orthogonal, and orthogonal to the code subspace.  Consider a damping on the $i^{th}$ qubit, where $i$ and $j$ are a pair.  The resulting state is stabilized by $Z_i$, $-Z_iZ_j,$ and the remaining $Z$-pair generators.  We will call this the $i^{th}$ damped subspace.  This subspace is clearly orthogonal to the code subspace due to the presence of the $-Z_iZ_j$ stabilizer. For the same reason, the $i^{th}$ damped subspace is clearly orthogonal to the $k^{th}$ damped subspace for $k\neq j$.  Finally, the $i^{th}$ and $j^{th}$ damped subspaces are orthogonal as we see that $Z_i$ stabilizes the $i^{th}$ and $-Z_i$ stabilizes the $j^{th}$.

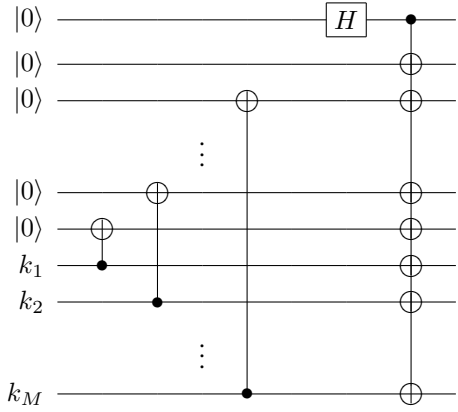
\begin{figure}[tb]
  \centerline{
    \Qcircuit @C=1.3em @R=.7em {
    \lstick{\ket{0}} & \qw & \qw & \qw & \qw & \qw & \gate{H} & \ctrl{13} & \qw \\
    \lstick{\ket{0}} & \qw & \qw & \qw & \qw & \qw & \qw & \targ & \qw \\
%    & & &  & & & & &\\
%    & & & \vdots & & & & &\\
%    & & &  & & & & &\\
%    \lstick{\ket{0}} & \qw & \qw & \qw & \qw & \qw & \qw & \targ & \qw \\
    \lstick{\ket{0}} & \qw & \qw & \qw & \targ & \qw & \qw & \targ & \qw \\
    & & &  & & & & &\\
    & & & \vdots & & & & &\\
    & & &  & & & & &\\
    \lstick{\ket{0}} & \qw & \targ & \qw & \qw & \qw & \qw & \targ & \qw \\
    \lstick{\ket{0}} & \targ & \qw & \qw & \qw & \qw & \qw & \targ & \qw \\
    \lstick{k_1} & \ctrl{-1} & \qw & \qw & \qw & \qw & \qw & \targ & \qw \\
    \lstick{k_2} & \qw & \ctrl{-3} & \qw & \qw & \qw & \qw & \targ & \qw \\
    & & &  & & & & &\\
    & & & \vdots & & & & &\\
    & & &  & & & & &\\
    \lstick{k_M} & \qw & \qw & \qw & \ctrl{-11} & \qw & \qw & \targ & \qw
    }
  }
  \caption{Circuit to encode the arbitrary state of $M$ qubits given in qubits $k_1\cdots k_M$ into $2(M+1)$ physical qubits.  This is the $[2(M+1),M]$ code in standard form.}\label{fig:generalized_leung_encoding}
\end{figure}

By writing the $[2(M+1),M]$ codes in the standard form, it is easy to generate an encoding circuit.  The circuit to encode the arbitrary state $\ket{\psi}$ in the $M$ qubits $k_1\cdots k_M$ is given in Fig.~\ref{fig:generalized_leung_encoding}.  The encoding circuit requires $3M+1$ CNOT operations and one Hadamard gate.

Let's write out the logical codewords  of the $[6,2]$ code given the choice of $\bar{Z}_i$ in Table \ref{tab:six-two and eight-three}:
\begin{eqnarray}
  \ket{00_L}&=&\frac{1}{\sqrt{2}}(\ket{000000}+\ket{111111})\\
  \ket{01_L}&=&\frac{1}{\sqrt{2}}(\ket{001001}+\ket{110110})\\
  \ket{10_L}&=&\frac{1}{\sqrt{2}}(\ket{000110}+\ket{111001})\\
  \ket{11_L}&=&\frac{1}{\sqrt{2}}(\ket{110000}+\ket{001111}).
\end{eqnarray}
Each codeword is the equal superposition of two basis states.  We can see by inspection that the damped subspaces are mutually orthogonal:  $E_1^{(k)}$ will eliminate one of the two basis states from each codeword and the resulting basis states do not overlap.

\subsection{Syndrome measurement}

We begin the recovery by first measuring the $Z$-pair stabilizers.  A $-1$ result on the $(i,j)$-pair stabilizer indicates a damping of either the $i^{th}$ or $j^{th}$ qubit.  This holds true even if multiple $Z$-pair stabilizers measure $-1$. Such a result indicates multiple damped qubits.  Once we have identified the qubit pair, we perform an additional stabilizer measurement to determine which of the qubits was damped.  As an example, if the $(i,j)$-pair was damped, we measure $Z_i$, with a $+1$ result indicating a damping on the $i^{th}$ qubit and a $-1$ indicating a damping on the $j^{th}$ qubit.  We perform this measurement for all pairs which measure $-1$.

%\begin{table}
%%\begin{center}
%%  \begin{tabular}{c}
%%    $[6,2]$ code\\
%%    \hline
%    \begin{tabular}{c@{}c@{}c@{}c@{}c@{}c}
%      X&X&X&X&X&X\\
%      Z&Z&I&I&I&I\\
%      I&I&Z&Z&I&I\\
%      I&I&I&I&Z&Z\\
%      \hline
%      X&X&I&I&I&I
%    \end{tabular}
%    \end{table}

%    $XXXXXX$\\
%    $ZZIIII$\\
%    $IIZZII$\\
%    $IIIIZZ$\\
%    \hline
%    $\bar{X_1}=XXIIII$\\
%    $\bar{X_2}=IIXXII$\\
%    $\bar{Z_1}=ZIIIZI$\\
%    $\bar{Z_2}=IIZIZI$
%  \end{tabular}
%  \hspace{20pt}
%  \begin{tabular}{c}
%    $[8,3]$ code\\
%    \hline
%    $XXXXXXXX$\\
%    $ZZIIIIII$\\
%    $IIZZIIII$\\
%    $IIIIZZII$\\
%    $IIIIIIZZ$\\
%    \hline
%    $\bar{X_1}=XXIIIIII$\\
%    $\bar{X_2}=IIXXIIII$\\
%    $\bar{X_3}=IIIIXXII$\\
%    $\bar{Z_1}=ZIIIIIZI$\\
%    $\bar{Z_2}=IIZIIIZI$\\
%    $\bar{Z_3}=IIIIZIZI$
%  \end{tabular}
%  \hspace{20pt}
%  \begin{tabular}{c}
%    $[10,4]$ code\\
%    \hline
%    $XXXXXXXXXX$\\
%    $ZZIIIIIIII$\\
%    $IIZZIIIIII$\\
%    $IIIIZZIIII$\\
%    $IIIIIIZZII$\\
%    $IIIIIIIIZZ$\\
%    \hline
%    $\bar{X_1}=XXIIIIIIII$\\
%    $\bar{X_2}=IIXXIIIIII$\\
%    $\bar{X_3}=IIIIXXIIII$\\
%    $\bar{X_4}=IIIIIIXXII$\\
%    $\bar{Z_1}=ZIIIIIIIZI$\\
%    $\bar{Z_2}=IIZIIIIIZI$\\
%    $\bar{Z_3}=IIIIZIIIZI$\\
%    $\bar{Z_4}=IIIIIIZIZI$
%  \end{tabular}
%  \end{center}
%  \caption[Stabilizers for six, eight, and ten qubit amplitude damping codes.]{Stabilizers for $[6,2]$, $[8,3]$, and $[10,4]$ qubit amplitude damping codes. These are generalized in a straightforward way from the $[4,1]$ code.}\label{tab:six-two and eight-three}
%\end{table}

\begin{table}
\begin{center}
\begin{tabular}{ccc} % Table will have 3 columns for the three codes and two rows, the first for the original form and the second for the standard form.
(A)&
  \begin{tabular}{c}
    $[6,2]$ code\\
    \hline
    \begin{tabular}{c@{}c@{}c@{}c@{}c@{}c}
    $X$&$X$&$X$&$X$&$X$&$X$\\
    $Z$&$Z$&$I$&$I$&$I$&$I$\\
    $I$&$I$&$Z$&$Z$&$I$&$I$\\
    $I$&$I$&$I$&$I$&$Z$&$Z$
    \end{tabular}
  \end{tabular}
  &
  \begin{tabular}{c}
    $[8,3]$ code\\
    \hline
    \begin{tabular}{c@{}c@{}c@{}c@{}c@{}c@{}c@{}c}
    $X$&$X$&$X$&$X$&$X$&$X$&$X$&$X$\\
    $Z$&$Z$&$I$&$I$&$I$&$I$&$I$&$I$\\
    $I$&$I$&$Z$&$Z$&$I$&$I$&$I$&$I$\\
    $I$&$I$&$I$&$I$&$Z$&$Z$&$I$&$I$\\
    $I$&$I$&$I$&$I$&$I$&$I$&$Z$&$Z$
    \end{tabular}
  \end{tabular}
%  &
%  \begin{tabular}{c}
%    $[10,4]$ code\\
%    \hline
%    \begin{tabular}{c@{}c@{}c@{}c@{}c@{}c@{}c@{}c@{}c@{}c}
%    $X$&$X$&$X$&$X$&$X$&$X$&$X$&$X$&$X$&$X$\\
%    $Z$&$Z$&$I$&$I$&$I$&$I$&$I$&$I$&$I$&$I$\\
%    $I$&$I$&$Z$&$Z$&$I$&$I$&$I$&$I$&$I$&$I$\\
%    $I$&$I$&$I$&$I$&$Z$&$Z$&$I$&$I$&$I$&$I$\\
%    $I$&$I$&$I$&$I$&$I$&$I$&$Z$&$Z$&$I$&$I$\\
%    $I$&$I$&$I$&$I$&$I$&$I$&$I$&$I$&$Z$&$Z$
%    \end{tabular}
%  \end{tabular}
  \\ % This is the line break for the big table.
%  & (A)  \\
  \\
  \hline
  \\
  (B)&
  
  \begin{tabular}{c}
    $[6,2]$ standard form\\
    \hline
    \begin{tabular}{c@{}c@{}c@{}c@{}c@{}c}
    $X$&$X$&$X$&$X$&$X$&$X$\\
    $Z$&$Z$&$I$&$I$&$I$&$I$\\
    $I$&$I$&$Z$&$I$&$I$&$Z$\\
    $I$&$I$&$I$&$Z$&$Z$&$I$
    \end{tabular}\\
    \begin{tabular}{c@{ = }c@{}c@{}c@{}c@{}c@{}c}
    $\bar{X_1}$&$I$&$I$&$I$&$X$&$X$&$I$\\
    $\bar{X_2}$&$I$&$I$&$X$&$I$&$I$&$X$\\
    $\bar{Z_1}$&$Z$&$I$&$I$&$I$&$Z$&$I$\\
    $\bar{Z_2}$&$Z$&$I$&$I$&$I$&$I$&$Z$
    \end{tabular}
 \end{tabular}
  &
  \begin{tabular}{c}
    $[8,3]$ standard form\\
    \hline
    \begin{tabular}{c@{}c@{}c@{}c@{}c@{}c@{}c@{}c}
    $X$&$X$&$X$&$X$&$X$&$X$&$X$&$X$\\
    $Z$&$Z$&$I$&$I$&$I$&$I$&$I$&$I$\\
    $I$&$I$&$Z$&$I$&$I$&$I$&$I$&$Z$\\
    $I$&$I$&$I$&$Z$&$I$&$I$&$Z$&$I$\\
    $I$&$I$&$I$&$I$&$Z$&$Z$&$I$&$I$\\
    \end{tabular}\\
    \begin{tabular}{c@{ = }c@{}c@{}c@{}c@{}c@{}c@{}c@{}c}
    $\bar{X_1}$&$I$&$I$&$I$&$I$&$X$&$X$&$I$&$I$\\
    $\bar{X_2}$&$I$&$I$&$I$&$X$&$I$&$I$&$X$&$I$\\
    $\bar{X_3}$&$I$&$I$&$X$&$I$&$I$&$I$&$I$&$X$\\
    $\bar{Z_1}$&$Z$&$I$&$I$&$I$&$I$&$Z$&$I$&$I$\\
    $\bar{Z_2}$&$Z$&$I$&$I$&$I$&$I$&$I$&$Z$&$I$\\
    $\bar{Z_3}$&$Z$&$I$&$I$&$I$&$I$&$I$&$I$&$Z$
    \end{tabular}
  \end{tabular}
%  &
%    \begin{tabular}{c}
%    $[10,4]$ standard form\\
%    \hline
%    \begin{tabular}{c@{}c@{}c@{}c@{}c@{}c@{}c@{}c@{}c@{}c}
%    $X$&$X$&$X$&$X$&$X$&$X$&$X$&$X$&$X$&$X$\\
%    $Z$&$Z$&$I$&$I$&$I$&$I$&$I$&$I$&$I$&$I$\\
%    $I$&$I$&$Z$&$I$&$I$&$I$&$I$&$I$&$I$&$Z$\\
%    $I$&$I$&$I$&$Z$&$I$&$I$&$I$&$I$&$Z$&$I$\\
%    $I$&$I$&$I$&$I$&$Z$&$I$&$I$&$Z$&$I$&$I$\\
%    $I$&$I$&$I$&$I$&$I$&$Z$&$Z$&$I$&$I$&$I$\\
%    \end{tabular}\\
%    \begin{tabular}{c@{ = }c@{}c@{}c@{}c@{}c@{}c@{}c@{}c@{}c@{}c}
%    $\bar{X_1}$&$I$&$I$&$I$&$I$&$I$&$X$&$X$&$I$&$I$&$I$\\
%    $\bar{X_2}$&$I$&$I$&$I$&$I$&$X$&$I$&$I$&$X$&$I$&$I$\\
%    $\bar{X_3}$&$I$&$I$&$I$&$X$&$I$&$I$&$I$&$I$&$X$&$I$\\
%    $\bar{X_4}$&$I$&$I$&$X$&$I$&$I$&$I$&$I$&$I$&$I$&$X$\\
%    $\bar{Z_1}$&$Z$&$I$&$I$&$I$&$I$&$I$&$Z$&$I$&$I$&$I$\\
%    $\bar{Z_2}$&$Z$&$I$&$I$&$I$&$I$&$I$&$I$&$Z$&$I$&$I$\\
%    $\bar{Z_3}$&$Z$&$I$&$I$&$I$&$I$&$I$&$I$&$I$&$Z$&$I$\\
%    $\bar{Z_4}$&$Z$&$I$&$I$&$I$&$I$&$I$&$I$&$I$&$I$&$Z$
%    \end{tabular}
%  \end{tabular}
%  \\
%  & (B) 
\end{tabular}
  \end{center}
  \caption[Stabilizers for six, eight, and ten qubit amplitude damping codes.]{Stabilizers for $[6,2]$ and $[8,3]$ amplitude damping codes. In (A), these are written in a way to illustrate the connection to the $[4,1]$ code.  In (B), we present the code in the standard form, which we achieve merely by swapping the code qubits and choosing the logical operators systematically.  The standard form provides a convenient description for generating quantum circuits for encoding.}\label{tab:six-two and eight-three}
\end{table}

If multiple stabilizers yield a $-1$ measurement then we have multiple damped qubits.  As before, this reduces by half the dimension of the subspace and we cannot preserve all logical superpositions.  For an example, examine the stabilizers for the $[6,2]$ code when both the first and fifth qubits are damped:
\begin{equation}
%  \begin{tabular}{c}
    \langle -ZZIIII,
    IIZIIZ,
    -IIIZZI,
    ZIIIII,
    IIIIZI\rangle.
%  \end{tabular}.
\end{equation}
This subspace has 5 stabilizers and thus has rank 2.  Furthermore, combining the last two stabilizers, we can see that $ZIIIZI=\bar{Z_1}$ stabilizes the subspace,  indicating that the remaining logical information is spanned by $\{\ket{01_L},\ket{00_L}\}$.  In general, for a $[2(M+1),M]$ code, up to $M+1$ dampings can be partially corrected as long as the dampings occur on distinct qubit pairs.  If $m$ is the number of damped qubits, then the resulting subspace has dimension $2^{M+1-m}$.

If all $Z$-pair measurements for the $[2(M+1),M]$ code return $+1$, we determine that we are in the `no dampings' syndrome and may perform some  further operation to reduce distortion as much as possible.  As in the example of the $[4,1]$ code in Sec.~\ref{sec:4qubit no damping}, we can choose to optimize this recovery with a $\gamma$-dependent recovery or we can apply a stabilizer projective measurement.  In the former case, we may calculate an optimized recovery with a SDP or any of the near-optimal methods of \cite{Fle:07,FleShoWin:J07a}.  If we choose a stabilizer measurement, we simply measure the all-$X$ generator $X\cdots X$ where a $+1$ result is a projection onto the code subspace.  A $-1$ result can be corrected by applying a $Z_i$ operation (in fact a $Z$ on any one of the qubits will suffice).  This can be seen by noting that the $-X\cdots X$ stabilizer changes the logical codewords by replacing the $+$ with a $-$.

\subsection{Stabilizer syndrome recovery operations}

In the previous section, we described syndrome measurements to determine which qubits were damped.  We also explained the extent to which multiple qubit dampings are correctable.  We now present a straightforward set of Clifford group operations to recover from each syndrome.

\begin{figure}
\centerline{
  \begin{tabular}{c}
    \Qcircuit @C=1.0em @R=1.0em {
    \lstick{\ket{0}} & \targ & \targ & \qw & \qw &  \qw &\qw & \qw & \qw & \qw &  \meter\\
    \lstick{\ket{0}} & \qw & \qw &\targ & \targ &  \qw & \qw & \qw & \qw & \qw &  \meter\\
    \lstick{\ket{0}} & \qw & \qw & \qw & \qw & \targ & \targ & \qw & \qw & \qw &  \meter\\
    & & & & & & & & & & \\
    & & & & & & & \vdots & \\
    & & & & & & & & & & \\
    \lstick{\ket{0}} & \qw & \qw & \qw & \qw & \qw & \qw & \qw & \targ & \targ &  \meter\\
    \lstick{k_1} & \ctrl{-7} & \qw & \qw & \qw & \qw & \qw & \qw & \qw & \qw & \qw \\
    \lstick{k_2}& \qw & \ctrl{-8} & \qw & \qw & \qw & \qw & \qw & \qw & \qw & \qw \\
    \lstick{k_3}& \qw & \qw & \ctrl{-8} & \qw & \qw & \qw & \qw & \qw & \qw & \qw \\
    \lstick{k_4}& \qw & \qw & \qw & \qw & \ctrl{-8} & \qw & \qw & \qw & \qw & \qw \\
    & & & & & & & & & & \\
    & & & & & & & \vdots & \\
    & & & & & & & & & & \\
    \lstick{k_{M+1}}& \qw & \qw & \qw & \qw & \qw & \qw & \qw & \ctrl{-8} & \qw & \qw \\
    \lstick{k_{M+2}}& \qw & \qw & \qw & \qw & \qw & \qw & \qw & \qw & \ctrl{-9} & \qw \\
    & & & & & & & & & & \\
    & & & & & & & \vdots \\
    & & & & & & & & & & \\
    \lstick{k_{2M+1}}& \qw & \qw & \qw & \qw & \qw & \ctrl{-17} & \qw & \qw & \qw & \qw \\
    \lstick{k_{2M+2}}& \qw & \qw & \qw & \ctrl{-19} & \qw & \qw & \qw & \qw & \qw & \qw
    }
    \\
    \\
    (A)\\
    \\
    \begin{tabular}{c@{\hspace{1cm}}c}
    \Qcircuit @C=1em @R=1em {
    \lstick{\ket{0}} & \gate{H} & \ctrl{6} & \gate{H} & \meter\\
    \lstick{k_1} & \qw & \targ & \qw & \qw \\
    \lstick{k_1} & \qw & \targ & \qw & \qw \\
    & &&&\\
    & \vdots &&&\\
    &&&&\\
    \lstick{k_{2M+2}} & \qw & \targ & \qw & \qw
    }&
    \Qcircuit @C=1em @R=1em {
    \lstick{\ket{0}} & \qw &\targ & \meter\\
    &&&\\
    &\vdots &&\\
    &&&\\
    \lstick{k_i} & \qw &\ctrl{-4} &\qw
    }\\
    &\\
    (B)&(C)
    \end{tabular}
    \end{tabular}
    }
  \caption{Syndrome measurement circuits for the $[2(M+1),M]$ code.  Circuit (A) measures each of the $Z$-pair stabilizers. If all of the measurements in (A) are $+1$, we are in the `no damping' syndrome and we perform the syndrome measurement in (B).  If the $(i,j)$-pair stabilizer measures $-1$, we perform the syndrome measurement in (C).}\label{fig:62syndrome}
\end{figure}
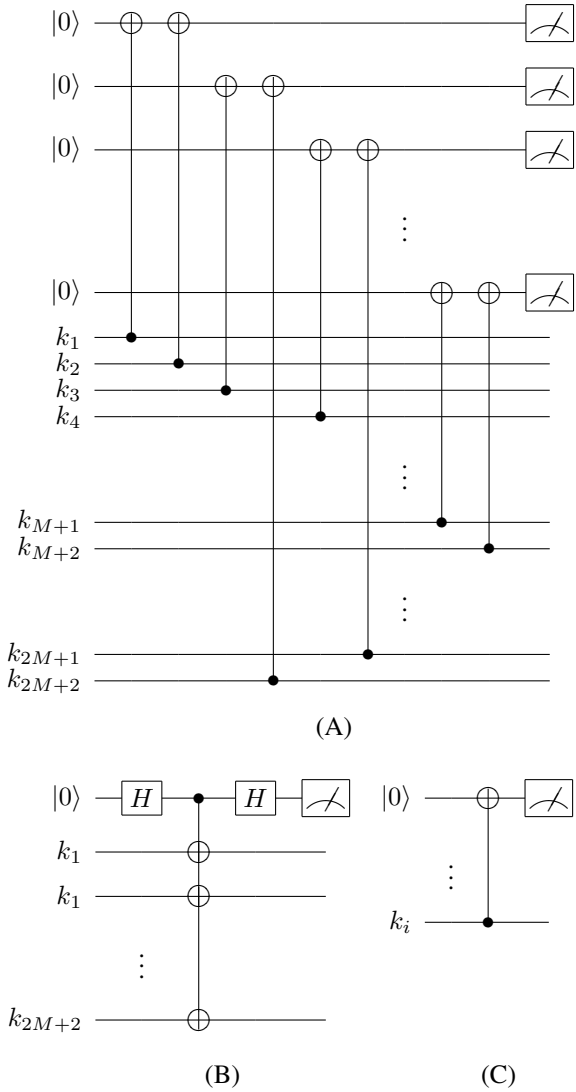

\begin{figure}
  \centerline{
    \Qcircuit @C=1.3em @R=.7em {
     & \gate{H} & \ctrl{5} & \gate{X} &\qw\\
     & \qw & \targ & \qw & \qw\\
     & \qw & \targ & \qw & \qw\\
     & \qw & \targ & \qw & \qw\\
     & \qw & \targ & \gate{X} & \qw\\
     & \qw & \targ & \qw & \qw\\
    }
  }
  \caption{Syndrome recovery circuit for the [6,2] code with the first and fifth qubits damped.}\label{fig:62syndrom recovery}
\end{figure}
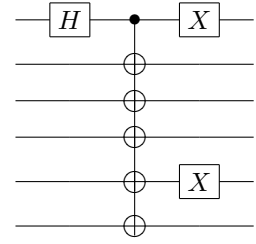

Consider a syndrome measurement in which we determine that $m$ qubits $i_1,\ldots,i_m$ were damped, where $m\leq M+1$.  We recover from this syndrome via the following three steps:
\begin{enumerate}
  \item{Apply a Hadamard gate $H_{i_1}$ on the $i_1$ qubit.}
  \item{With qubit $i_1$ as the control, apply a CNOT gate to every other qubit.}
  \item{Flip every damped qubit: $X_{i_1}\cdots X_{i_m}$.}
\end{enumerate}
The procedure is illustrated as a quantum circuit for a two-damping syndrome and the $[6,2]$ code in Fig.~\ref{fig:62syndrom recovery}.

To see that this is the correct syndrome recovery for the $[2(M+1),M]$ code, we need to examine the effect of the three gate operations on the damped subspace stabilizers.  In the syndrome where $i_1,\ldots,i_m$ are damped, we have three categories of generators for the resulting stabilizer group: $-Z$-pair stabilizers for the damped pairs, $+Z$-pair stabilizers for the non-damped pairs, and $Z_{i_1},\ldots,Z_{i_m}$ for each damped qubit.  We need to see the effect of the recovery gate operations on each of these generators.  Fortunately, we can demonstrate all of the relevant cases with the example of the $[6,2]$ code with the first and fifth qubits damped:
\begin{eqnarray}
  \begin{tabular}{c@{}c@{}c@{}c@{}c@{}c}
    -Z&Z&I&I&I&I\\
    I&I&Z&I&I&Z\\
    -I&I&I&Z&Z&I\\
    Z&I&I&I&I&I\\
    I&I&I&I&Z&I
  \end{tabular}
  &\rightarrow^{H_1}&
  \begin{tabular}{c@{}c@{}c@{}c@{}c@{}c}
    -X&Z&I&I&I&I\\
    I&I&Z&I&I&Z\\
    -I&I&I&Z&Z&I\\
    X&I&I&I&I&I\\
    I&I&I&I&Z&I
  \end{tabular}
  \rightarrow^{\textrm{CNOT}_1\textrm{'s}}
  \begin{tabular}{c@{}c@{}c@{}c@{}c@{}c}
    -Y&Y&X&X&X&X\\
    I&I&Z&I&I&Z\\
    -I&I&I&Z&Z&I\\
    X&X&X&X&X&X\\
    Z&I&I&I&Z&I
  \end{tabular}
  \nonumber\\
  &\rightarrow^{X_1X_5}&
  \begin{tabular}{c@{}c@{}c@{}c@{}c@{}c}
    Y&Y&X&X&X&X\\
    I&I&Z&I&I&Z\\
    I&I&I&Z&Z&I\\
    X&X&X&X&X&X\\
    Z&I&I&I&Z&I
  \end{tabular}
  =
  \begin{tabular}{c@{}c@{}c@{}c@{}c@{}c}
    Z&Z&I&I&I&I\\
    I&I&Z&I&I&Z\\
    I&I&I&Z&Z&I\\
    X&X&X&X&X&X\\
    Z&I&I&I&Z&I
  \end{tabular}.
\end{eqnarray}
The final two sets of stabilizers are equivalent since $ZZIIII$ is the product of $XXXXXX$ and $YYXXXX$.  The first four generators of the resulting group are the code stabilizer.  The last generator is $\bar{Z}_1$ which, as we saw before, indicates that the recovered information is spanned by $\{\ket{00_L},\ket{01_L}\}$ while the other two dimensions of information have been lost.

While we have shown that the syndrome recovery operation returns the information to the code subspace, it remains to demonstrate that the information is correctly decoded.  We can demonstrate this by considering the syndrome recovery operation on each of the $\bar{Z}_i$ of the code.  By showing that each of these is correctly preserved, we conclude that the syndrome recovery operation is correct.

We have chosen the $\bar{Z}_i$ so that each has exactly two qubit locations with a $Z$ while the rest are $I$.  There are, therefore, five cases of interest.  In case 1, neither of the damped qubits corresponds to a location with a $Z$.  In case 2, the first damped qubit ($i_1$) corresponds to a location with a $Z$.  In case 3, one of the $Z$ locations corresponds to a damped qubit, but it is not $i_1$.  In case 4, both of the $Z$ locations correspond to a damped qubit, but neither is $i_1$.
Finally, case 5 is when both $Z$ locations correspond to damped qubits and one is $i_1$.

Without loss of generality, we can see the effect of each case by considering an example using $ZIIIZI$ and appropriately selected damped qubits.  Consider case 1 where we can let $i_1=2$:
\begin{eqnarray}
  ZIIIZI&\rightarrow^{H_{i_1}}&ZIIIZI\rightarrow^{\textrm{CNOT}_{i_1}\textrm{'s}}ZIIIZI\nonumber\\
  &\rightarrow^{X_{i_1}\cdots X_{i_m}}& ZIIIZI.
\end{eqnarray}
In case 2, $i_1=1$:
\begin{eqnarray}
  -ZIIIZI&\rightarrow^{H_{i_1}}&-XIIIZI\rightarrow^{\textrm{CNOT}_{i_1}\textrm{'s}}-YXXXYX\nonumber\\
  &\rightarrow^{X_{i_1}\cdots X_{i_m}}& YXXXYX.
\end{eqnarray}
Notice that this last is equivalent to $ZIIIZI$ as $XXXXXX$ is in the stabilizer.\\
In case 3, $i_1=2$ while $i_2=5$:
\begin{eqnarray}
  -ZIIIZI&\rightarrow^{H_{i_1}}&-ZIIIZI\rightarrow^{\textrm{CNOT}_{i_1}\textrm{'s}}-ZIIIZI\nonumber\\
  &\rightarrow^{X_{i_1}\cdots X_{i_m}}& ZIIIZI.
\end{eqnarray}
In case 4, let $i_1=2,$, $i_2=1$, and $i_3=5$:\footnote{While this contradicts our statement that the lowest numbered qubit would be $i_1$, the assignment of $i_1=2$ when the first qubit is also damped has no impact on the argument.}
\begin{eqnarray}
  ZIIIZI&\rightarrow^{H_{i_1}}&ZIIIZI\rightarrow^{\textrm{CNOT}_{i_1}\textrm{'s}}ZIIIZI\nonumber\\
  &\rightarrow^{X_{i_1}\cdots X_{i_m}}& ZIIIZI.
\end{eqnarray}
In case 5, $i_1=1$ and $i_2=5$:
\begin{eqnarray}
  ZIIIZI&\rightarrow^{H_{i_1}}&XIIIZI\rightarrow^{\textrm{CNOT}_{i_1}\textrm{'s}}YXXXYX\nonumber\\
  &\rightarrow^{X_{i_1}\cdots X_{i_m}}& YXXXYX.
\end{eqnarray}

We see that in all cases, the recovery procedure correctly preserves the geometry of the encoded information, even in the case of multiple qubit dampings.  It is worth emphasizing, however, that when multiple qubits are damped at least half of the information dimensions are lost.

\subsection{Performance comparison}

\begin{figure}
  \begin{center}
  \begin{tabular}{c}
  \includegraphics[width=.8\columnwidth]{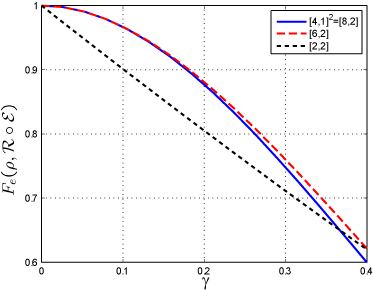}\\
  (A)\\
  \includegraphics[width=.8\columnwidth]{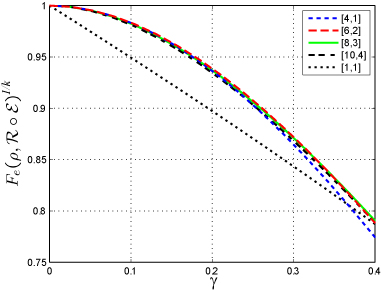}\\
  (B)
  \end{tabular}
    \caption[Performance comparison for generalized amplitude damping codes.]{Performance omparison of generalized amplitude damping codes.  In (A) we compare the $[6,2]$ code with the $[4,1]$ repeated twice.  In (B), we compare the $[4,1]$, $[6,2]$, $[8,3]$ and $[10,4]$ codes.  The entanglement fidelity has been normalized as $1/k$ where $k$ is the number of encoded qubits.  Notice that despite the increasing rates, the normalized entanglement fidelity maintains high performance.}\label{fig:Generalized_Leung}
  \end{center}
\end{figure}

It is useful to compare the performance of each of the $[2(M+1),M]$ codes in terms of the damping parameter $\gamma$.  Consider a comparison between the $[4,1]$ code and the $[6,2]$ code.  To make a valid comparison, we need to establish a common baseline.  We do this by considering the encoding of two qubits with the $[4,1]$ code. For the completely mixed state $\rho=I/2$, this is the equivalent of squaring the single qubit entanglement fidelity:
\begin{equation}
  \bar{F}_e(\rho\otimes\rho,\R\circ\E\otimes\R\circ\E)=\bar{F}_e(\rho,\R\circ\E)^2.
\end{equation}
This comparison is given in Fig.~\ref{fig:Generalized_Leung} (A).  To compare multiple codes, it is more straightforward to normalize each to a single qubit baseline.  This can be done by computing $\bar{F}_e^{(1/k)}$ for an $[n,k]$ code.  The normalized performance for the $[4,1]$, $[6,2]$, $[8,3]$ and $[10,4]$ codes is given in Fig.~\ref{fig:Generalized_Leung} (B).

It is very interesting to note how comparably these codes maintain the fidelity even as the code rate increases.  This is particularly striking when noting that each code can still perfectly correct only a single damping error.  Thus, the $[4,1]^{\otimes 4}$ can correct 4 dampings (as long as they occur on separate blocks) while the $[10,4]$ code can only perfectly correct 1.  Yet we see that the normalized performance is quite comparable.

\begin{figure}
  \begin{center}
    \includegraphics[width=\columnwidth]{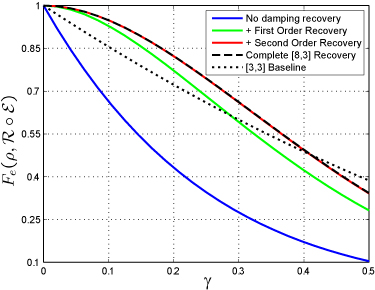}
    \caption[Fidelity contributions for each order error of the eight qubit amplitude damping code.]{Fidelity contributions for each order error of the $[8,3]$ amplitude damping code.  We see that the no damping, first, and second order recovery syndromes contribute to the entanglement fidelity of the recovery operation.}\label{fig:83fidelity_contributions}
  \end{center}
\end{figure}

We take a closer look at the performance of the $[8,3]$ code in Fig.~\ref{fig:83fidelity_contributions}.  We see that, while most of the entanglement fidelity is supplied by correcting no damping and $E_1^{(i)}$ terms, a not insignificant performance benefit arises by partially correcting second order damping errors.  In the case of the $[4,1]$ recovery, we concluded that such contributions improved the entanglement fidelity, but not the minimum fidelity as $\ket{1_L}$ was never preserved by such a recovery.  This is not the case for the higher rates.  Two damping errors eliminate half of the logical space, but different combinations of damping errors will divide the logical space differently.  For example, an damping error on the fifth and sixth qubits means the resulting space is stabilized by $\bar{Z}_1\bar{Z}_2$ thus eliminating logical states $\ket{01x_L}$ and $\ket{10x_L}$ (where $x$ indicates either $0$ or $1$).  On the other hand, a damping on the fifth and seventh qubits results in a space stabilized by $\bar{Z}_1\bar{Z}_3$ eliminating logical states $\ket{0x1_L}$ and $\ket{1x0_L}$.  Thus, correcting second order damping errors still contributes to minimum fidelity performance.

\begin{table}
  \begin{center}
    \begin{tabular}{c}
        Gottesman $[8,3]$ code\\
        \hline
        \begin{tabular}{c@{}c@{}c@{}c@{}c@{}c@{}c@{}c}
        $X$&$X$&$X$&$X$&$X$&$X$&$X$&$X$\\
        $Z$&$Z$&$Z$&$Z$&$Z$&$Z$&$Z$&$Z$\\
        $I$&$X$&$I$&$X$&$Y$&$Z$&$Y$&$Z$\\
        $I$&$X$&$Z$&$Y$&$I$&$X$&$Z$&$Y$\\
        $I$&$Y$&$X$&$Z$&$X$&$Z$&$I$&$Y$
        \end{tabular}
    \end{tabular}
  \end{center}
  \caption{Stabilizers for the [8,3] code due to Gottesman\cite{Got:97}.}\label{tab:Gottesman 83}
\end{table}
Given their identical rates, it is reasonable to compare the $[8,3]$ amplitude damping code presented here with the generic $[8,3]$ stabilizer code due to Gottesman\cite{Got:97}.  The stabilizers for this code are presented in Table \ref{tab:Gottesman 83}.  This code can correct an arbitrary single qubit error, and thus can correct all first order amplitude damping errors, as well as the less probable $Z$ errors.  These are corrected with 25 stabilizer syndrome measurements (Pauli operators on each of the 8 qubits as well as the identity).  This leaves an additional 7 degrees of freedom to correct for higher order errors.  While typically these are not specified, since we know the channel of interest is the amplitude damping channel, we can do a small amount of channel-adaptation by selecting appropriate recovery operations for these syndromes.  Since $X$ and $Y$ errors are the most common, we choose operators with 2 $X$'s or 2 $Y$'s (or one of each).

\begin{figure}
  \begin{center}
    \includegraphics[width=\columnwidth]{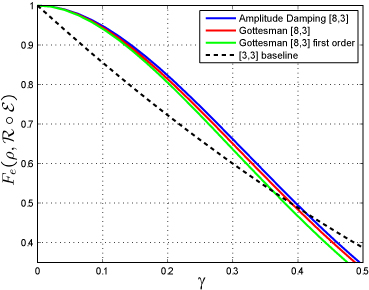}
    \caption[Comparison of the amplitude damping rate $3/8$ code and the generic rate $3/8$ code due to Gottesman.]{Comparison of the amplitude damping $[8,3]$ code and the generic rate $[8,3]$ code due to Gottesman.  We include both the Gottesman recovery where no attention is paid to second order recoveries, as well as a recovery where second order syndromes are chosen to adapt to the amplitude damping channel.}\label{fig:83Gottesman_compare}
  \end{center}
\end{figure}

The comparison between the rate $3/8$ codes is given Fig.~\ref{fig:83Gottesman_compare}.  Here we see that the channel-adapted $[8,3]$ code outperforms the generic Gottesman code, but the effect is minor.  The attention to higher order syndromes is seen to improve the performance of the $[8,3]$ code modestly.  It should be pointed out that both recovery operations can be accomplished with Clifford group operations, and neither is dependent on $\gamma$.

\section{Linear codes for the amplitude damping channel}

The channel-adapted codes of the previous section have similar corrective properties to the $[4,1]$ code: $\{I,E_1^{(i)}\}$ are correctable errors while $\{X_i,Y_i\}$ are not.  It is actually quite simple to design channel-adapted codes that correct both $X_i$ and $Y_i$ errors and thus can correct $\{I,E_1^{(i)}\}$ as well. Consider the $[7,3]$ code presented in Table \ref{table:seven three}.  The first three stabilizers can be readily identified as the classical $[7,4]$ Hamming code parity check matrix (replacing 0 with $I$ and 1 with $Z$).  They are also three of the six stabilizers for the Steane code.  Measuring these three stabilizers, an $X_i$ will result in a unique three bit measurement syndrome $(M_1,M_2,M_3)$.  (In fact, a nice property of the Hamming code is that the syndrome, replacing $+1$ with 0 and $-1$ with 1, is just the binary representation of $i$, the qubit that sustained the error.)  Unfortunately, a $Y_i$ error will yield the same syndrome as $X_i$.  We add the $XXXXXXX$ generator to distinguish the two, resulting in 14 orthogonal error syndromes for the $\{X_i,Y_i\}_{i=1}^7$.

\begin{table}
  \begin{center}
    \begin{tabular}{c}
      $[7,3]$ linear code\\
      \hline
    \begin{tabular}{c@{}c@{}c@{}c@{}c@{}c@{}c}
      $I$&$I$&$I$&$Z$&$Z$&$Z$&$Z$\\
      $I$&$Z$&$Z$&$I$&$I$&$Z$&$Z$\\
      $Z$&$I$&$Z$&$I$&$Z$&$I$&$Z$\\
      $X$&$X$&$X$&$X$&$X$&$X$&$X$
      \end{tabular}
    \end{tabular}
  \end{center}
  \caption[Amplitude damping channel-adapted seven qubit linear code.]{Amplitude damping channel-adapted $[7,3]$ linear code. Looking at the first three generators, this is clearly based on the classical Hamming code. The fourth generator differentiates between $X$ and $Y$ syndromes.}\label{table:seven three}
\end{table}

\begin{figure}[tb]
  \centerline{
    \Qcircuit @C=1.0em @R=1.0em {
    \lstick{\ket{0}} & \targ & \qw & \targ & \targ & \qw & \qw & \targ & \qw & \qw & \qw &\meter\\
    \lstick{\ket{0}} & \qw & \targ & \targ & \qw & \targ & \qw & \targ & \qw & \qw & \qw &\meter\\
    \lstick{\ket{0}} & \targ & \targ & \targ & \qw & \qw & \targ & \qw & \qw & \qw & \qw &\meter\\
    \lstick{\ket{0}} & \qw & \qw & \qw & \qw & \qw & \qw & \qw & \gate{H} & \ctrl{7} & \gate{H} &\meter\\
    & \ctrl{-4} & \qw & \qw & \qw & \qw & \qw & \qw & \qw & \targ & \qw&\qw\\
    & \qw & \ctrl{-4} & \qw & \qw & \qw & \qw & \qw & \qw & \targ &\qw&\qw\\
    & \qw & \qw & \ctrl{-6} & \qw & \qw & \qw & \qw & \qw & \targ &\qw&\qw\\
    & \qw & \qw & \qw &\ctrl{-7} &  \qw & \qw & \qw & \qw & \targ &\qw&\qw\\
    & \qw & \qw & \qw & \qw &\ctrl{-7} &  \qw & \qw & \qw & \targ &\qw&\qw\\
    & \qw & \qw & \qw & \qw & \qw &\ctrl{-7} &  \qw & \qw & \targ &\qw&\qw\\
    & \qw & \qw & \qw & \qw & \qw & \qw &\ctrl{-10} &  \qw & \targ &\qw&\qw
    }
  }
  \caption{Syndrome measurement circuit for the $[7,3]$ amplitude damping code.}\label{fig:73syndrome}
\end{figure}
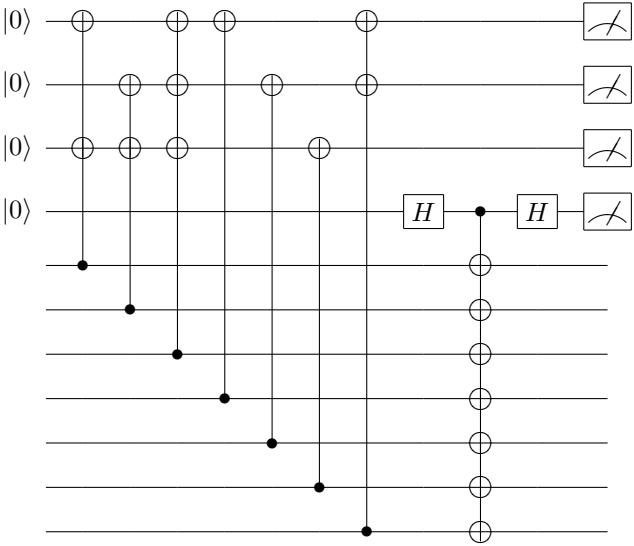

\begin{figure}
  \begin{center}
    \includegraphics[width=\columnwidth]{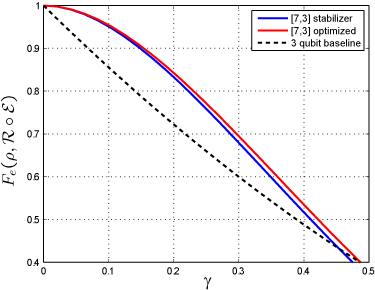}
    \caption[Optimal vs.~code projection recovery operations for the seven qubit linear amplitude damping code.]{Optimal vs.~code projection recovery operations for the [7,3] code.  We compare the entanglement fidelity for the optimal recovery operation and the recovery that includes a projection onto the code subspace.  For comparison, we also include the baseline performance of three unencoded qubits.  While the optimal recovery outperforms the code projector recovery, the performance benefit is likely small compared to the cost of implementing the optimal.}\label{fig:73optimized}
  \end{center}
\end{figure}

As in previous examples, we have a choice of recovery operations for the `no dampings' syndrome.  We can minimize the `no damping' distortion as was done in previous cases by computing the optimal or structured near-optimal recovery within this subspace.  This will result in a $\gamma$-dependent recovery operation.  Alternatively, we can simply measure $XXXXXXX$ with a $+1$ projecting onto the code subpsace and a $-1$ requiring a correction of $Z_i$.  We compare these recovery operations in Fig.~\ref{fig:73optimized}.

\begin{figure}
  \begin{center}
    \includegraphics[width=\columnwidth]{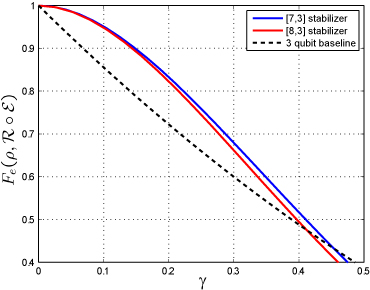}
    \caption[Comparison of the seven and eight qubit amplitude damping codes.]{Comparison of the $[7,3]$ and $[8,4]$ qubit amplitude damping codes.  We see that the $[7,3]$ performance is slightly better, despite the higher rate.}\label{fig:AmpDamp87}
  \end{center}
\end{figure}

We see in Fig.~\ref{fig:AmpDamp87} that the $[7,3]$ code slightly outperforms the $[8,3]$ code of Sec.~\ref{sec:generalized 41}.  The $[7,3]$ code perfectly corrects first order dampings and does not correct any second order dampings while the $[8,3]$ code partially corrects for higher order dampings.  The performance advantage of the $[7,3]$ code arises from the decreased length: the probability of a higher order damping error decreases as only seven physical qubits are needed.

Given its structure, it is logical to compare the $[7,3]$ amplitude damping code to the $[7,1]$ Steane code, as both are derived from the classical Hamming code.  It was shown in \cite{FleShoWin:J07a,Fle:07} that the Steane code is not particularly well adaptable to amplitude damping errors; despite its extra redundancy, the channel-adapted $[5,1]$ code significantly outperforms the channel-adapted Steane code.  This is particularly unfortunate as the Steane code can be implemented with such efficiency, with particular value for fault tolerant quantum computing.  The $[7,3]$ code provides a useful compromise position.

\begin{figure}
  \begin{center}
    \includegraphics[width=\columnwidth]{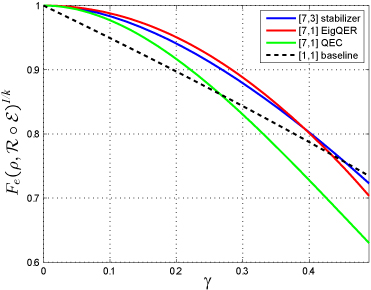}
    \caption[Comparison of the seven qubit amplitude damping code and the seven qubit Steane code.]{Comparison of the $[7,1]$ Steane code and the $[7,3]$ amplitude damping code, normalized by ${}^{1/k}$.    We see that the $[7,3]$ performance is very similar to the EigQER optimized recovery for the Steane code.}\label{fig:AmpDamp73 vs 71}
  \end{center}
\end{figure}

We see in Fig.~\ref{fig:AmpDamp73 vs 71} the performance comparison for $[7,3]$ code and the $[7,1]$ code (with and without channel-adapted recovery).   It comes as no surprise that the $[7,3]$ code outperforms the $[7,1]$ with standard stabilizer recovery: each perfectly corrects the first order damping errors, but the $[7,3]$ code has done so while preserving three times as much information.  It is interesting to see how close the $[7,3]$ performance is to the channel-adapted $[7,1]$.  It was shown in \cite{Fle:07,FleShoWin:J07a} that the channel-adapted $[7,1]$ at least partially corrects some second-order damping errors; the $[7,3]$ code does not.  This is mitigated by the higher rate of the $[7,3]$ code as again, three times as much information is preserved.

The $[7,3]$ code is not the only high rate linear code for amplitude damping errors.  Consider any classical linear code that can correct 1 error for which codewords have even parity.  We can convert this code to a quantum amplitude damping code in the same way as the $[7,3]$ code.  If $H$ is the parity check matrix for an $[n,k]$ classical linear code, then each row can be made a quantum code stabilizer replacing 1's with $Z$ and 0's with $I$.  To distinguish $X_i$ and $Y_i$ errors we include $X^{\otimes n}$ as a generator.  Since the classical code has even parity, we know that this generator commutes with the others.  This construction yields a $[n,k-1]$ quantum amplitude damping code that corrects for single amplitude damping errors.

The $[7,3]$ code we have presented here follows the structure proposed in \cite{Got:97} for amplitude damping codes; namely the code is a combination of an $X$-error correcting code and a $Z$-error detecting code.  It is not immediately clear how to generalize to $t$ error correcting linear codes.  Instead of a single generator to distinguish $X_i$ and $Y_i$ errors, we require an extra $t$ generators as we must distinguish $X_i$ and $Y_i$ for each corrected damping.

\section{Amplitude damping errors and the Shor code}

We now turn our attention to the $[9,1]$ Shor code and its performance with a recovery operation channel-adapted to amplitude damping errors.  The structured near-optimal results in \cite{Fle:07,FleShoWin:J07a} showed that the Shor code provides remarkably good protection from the amplitude damping channel.  Furthermore, the structured recovery operation is essentially optimal.  Thus, in this case, the optimal channel-adapted recovery operation can be described as a projective syndrome measurement followed by a unitary operation.   Given this intuitive structure, we can analyze the amplitude damping channel-adapted recovery operation.

We first note that first order errors $\{E_1^{(k)}\}$ are perfectly correctable.  This comes as no surprise, since the Shor code can correct an arbitrary single qubit operation.  What may be surprising is that second order errors $\{E_1^{(j)}E_1^{(k)}\}$ are also perfectly correctable.  This was pointed out in \cite{Got:97} and can be seen through the same kind of stabilizer analysis of damped subspaces as we employed for the $[2(M+1),M]$ codes.

\begin{table}
\begin{center}
\begin{tabular}{c}
  \begin{tabular}{cc}
     \begin{tabular}{c}
        Qubit 1 damped\\
        \hline
        \begin{tabular}{c@{}c@{}c@{}c@{}c@{}c@{}c@{}c@{}c@{}c}
            -&$Z$&$Z$&$ I$&$ I$&$ I$&$ I$&$ I$&$ I$&$ I$\\
            &$I $&$Z$&$ Z$&$ I$&$ I$&$ I$&$ I$&$ I$&$ I$\\
            &$I$&$ I$&$ I$&$ Z$&$ Z$&$ I$&$ I$&$ I$&$ I$\\
            &$I$&$ I$&$ I$&$ I$&$ Z$&$ Z$&$ I$&$ I$&$ I$\\
            &$I$&$ I$&$ I$&$ I$&$ I$&$ I$&$ Z$&$ Z$&$ I$\\
            &$I$&$ I$&$ I$&$ I$&$ I$&$ I$&$ I$&$ Z$&$ Z$\\
            &$I$&$ I$&$ I$&$ X$&$ X$&$ X$&$ X$&$ X$&$ X$\\
            &$Z$&$I$&$I$&$I$&$I$&$I$&$I$&$I$&$I$
        \end{tabular}
     \end{tabular}
     &
     \begin{tabular}{c}
        Qubits 2 \& 3 damped\\
        \hline
        \begin{tabular}{c@{}c@{}c@{}c@{}c@{}c@{}c@{}c@{}c@{}c}
            -&$Z$&$Z$&$ I$&$ I$&$ I$&$ I$&$ I$&$ I$&$ I$\\
            &$I $&$Z$&$ Z$&$ I$&$ I$&$ I$&$ I$&$ I$&$ I$\\
            &$I$&$ I$&$ I$&$ Z$&$ Z$&$ I$&$ I$&$ I$&$ I$\\
            &$I$&$ I$&$ I$&$ I$&$ Z$&$ Z$&$ I$&$ I$&$ I$\\
            &$I$&$ I$&$ I$&$ I$&$ I$&$ I$&$ Z$&$ Z$&$ I$\\
            &$I$&$ I$&$ I$&$ I$&$ I$&$ I$&$ I$&$ Z$&$ Z$\\
            &$I$&$ I$&$ I$&$ X$&$ X$&$ X$&$ X$&$ X$&$ X$\\
            -&$Z$&$I$&$I$&$I$&$I$&$I$&$I$&$I$&$I$
        \end{tabular}
     \end{tabular}
     \end{tabular}
     \\
     \\
     \begin{tabular}{c}
        Qubits 1 \& 7 damped\\
        \hline
        \begin{tabular}{c@{}c@{}c@{}c@{}c@{}c@{}c@{}c@{}c@{}c}
            -&$Z$&$Z$&$ I$&$ I$&$ I$&$ I$&$ I$&$ I$&$ I$\\
            &$I $&$Z$&$ Z$&$ I$&$ I$&$ I$&$ I$&$ I$&$ I$\\
            &$I$&$ I$&$ I$&$ Z$&$ Z$&$ I$&$ I$&$ I$&$ I$\\
            &$I$&$ I$&$ I$&$ I$&$ Z$&$ Z$&$ I$&$ I$&$ I$\\
            -&$I$&$ I$&$ I$&$ I$&$ I$&$ I$&$ Z$&$ Z$&$ I$\\
            &$I$&$ I$&$ I$&$ I$&$ I$&$ I$&$ I$&$ Z$&$ Z$\\
            &$Z$&$I$&$I$&$I$&$I$&$I$&$I$&$I$&$I$\\
            &$I$&$I$&$I$&$I$&$I$&$I$&$Z$&$I$&$I$
        \end{tabular}
     \end{tabular}
%     &
%     \begin{tabular}{c}
%        Qubit 1 damped\\
%        \hline
%        \begin{tabular}{c@{}c@{}c@{}c@{}c@{}c@{}c@{}c@{}c}
%            $Z$&$Z$&$ I$&$ I$&$ I$&$ I$&$ I$&$ I$&$ I$\\
%            $I $&$Z$&$ Z$&$ I$&$ I$&$ I$&$ I$&$ I$&$ I$\\
%            $I$&$ I$&$ I$&$ Z$&$ Z$&$ I$&$ I$&$ I$&$ I$\\
%            $I$&$ I$&$ I$&$ I$&$ Z$&$ Z$&$ I$&$ I$&$ I$\\
%            $I$&$ I$&$ I$&$ I$&$ I$&$ I$&$ Z$&$ Z$&$ I$\\
%            $I$&$ I$&$ I$&$ I$&$ I$&$ I$&$ I$&$ Z$&$ Z$\\
%            $X$&$ X$&$ X$&$ X$&$ X$&$ X$&$ I$&$ I$&$ I$\\
%            $I$&$ I$&$ I$&$ X$&$ X$&$ X$&$ X$&$ X$&$ X$
%        \end{tabular}
%     \end{tabular}
  \end{tabular}
  \end{center}
\caption{Stabilizers for several damped subspace syndromes for the Shor code.}\label{tab:Shor subspaces}
\end{table}

We see in Table \ref{tab:Shor subspaces} a few representative syndrome subspaces for damping errors on the Shor code.  From these subspaces, we surmise that the first step in making a syndrome measurement is to measure the first 6 code stabilizers (each of which has a pair of $Z$'s).  Depending on those outcomes, we can make a further stabilizer measurement.

As an example, consider when the first stabilizer returns a $-1$ and the rest return $+1$.  In that case, we can conclude that either the first qubit was damped, or both the second and third qubits were damped.  These can be distinguished by measuring $Z_1$ with a $+1$ indicating qubit one and a $-1$ indicating both qubits two and three.

It is interesting to note that in this case, we will have only measured 7 stabilizers, and thus need one further measurement to achieve a 2 dimensional subspace.  A natural choice would be to measure $IIIXXXXXX$, but this is an opportunity for a $\gamma$-dependent measurement instead.  As before, such an operation can improve performance at the cost of circuit complexity.  In most of this chapter, we have leaned toward the simpler operation, concluding that $\gamma$-dependent operations provide some performance benefit but not enough to justify the added complexity.  We will see that in the case of the Shor code, the performance gain may be sufficiently large to warrant a $\gamma$-dependent operation.

Before this consideration, let's turn to another syndrome for multiple qubit dampings.  We already examined an example where the second and third qubits are both damped.  The Shor code is divided into three blocks of three qubits each; this case extends exactly to all circumstances when both damped qubits fall on the same block.  The third example in Table \ref{tab:Shor subspaces} is an example of two qubits damped from different blocks; in this case the first and fifth stabilizers are both measured to be $-1$.  While the most likely cause of this syndrome is a two-qubit damping, we can further measure $Z_1$ and $Z_7$ to correct for a three or four-qubit damping occurrence.

\begin{figure}
  \includegraphics[width=\columnwidth]{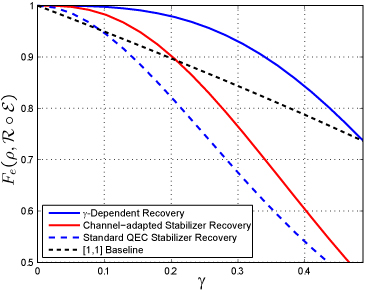}
  \caption{Channel-adapted stabilizer recovery vs.~$\gamma$-dependent recovery for the Shor code and the amplitude damping channel}\label{fig:Shor_recoveries}
\end{figure}

The preceding discussion of stabilizer subspaces provides two alternative recovery operations.  We may begin with a projective syndrome measurement of the first 6 code generators.  At that point, we may either make a set of stabilizer measurements to project onto the damped subspaces, or we may make a $\gamma$-dependent syndrome recovery to minimize this distortion.  It turns out that the best $\gamma$-dependent syndrome recovery has equivalent performance to the structured near-optimal recovery operation and is therefore essentially optimal.  The stabilizer recovery, while simple to implement with Clifford group operations, has significantly weaker performance.  We compare the two recovery operations for various values of $\gamma$ in Fig.~\ref{fig:Shor_recoveries}.

How should we understand the extensive performance gains for the $\gamma$-dependent recovery?  Both are consequences of the $E_0$ distortion imparted onto the quantum state and the degrees of freedom in the code.  The $\gamma$-dependent operation arises when we have a remaining degree of freedom after determining the syndrome.  For the $[2(M+1),M]$ codes and the $[7,3]$ code, we only have such freedom in the `no damping' syndrome; in all of the damping syndromes, the syndrome measurement requires a full set of stabilizer measurements.  We saw that for the Shor code first order dampings require only 7 stabilizer measurements to determine the syndrome, leaving one extra degree of freedom.  We also have an extra degree of freedom when two qubits from the same block are damped.  These constitute all of the first and some of the second order syndromes, each of which can be optimized to minimize $E_0$ distortion.

\section{Conclusion}

We have developed several quantum error correcting codes channel-adapted for the amplitude damping channel.  All of the encodings can be compactly described in the stabilizer formalism.  While optimized $\gamma$-dependent recovery operations are possible, a much simpler recovery operation using only stabilizer measurements and Clifford group operations achieves nearly equivalent performance.
The channel-adapted codes have much higher rates (with short block lengths) than generic quantum codes.

The creation of straightforward codes for the amplitude damping channel is a major step toward the still-open question of channel-adapted fault tolerant quantum computing.  Intuitively, channel-adaptation should be able to improve fault tolerant thresholds and reduce the necessary overhead for fault tolerance.  Before such intuition is confirmed for the amplitude damping channel, several obstacles must be overcome.  As an example, we must construct a universal set of operations on one or more of the channel-adapted codes presented here.  Such challenges are acknowledged, but deferred for future consideration.
\bibliographystyle{IEEEtran}
\bibliography{quantum}

\end{document}